\documentclass[12pt]{iopart}

\usepackage{array}
\usepackage{epsfig}
\usepackage{amssymb}

\def\0{\varnothing}

\begin{document}
\title[Ergodicity breaking in one-dimensional reaction-diffusion systems]{Ergodicity breaking in one-dimensional reaction-diffusion systems}
\author{A.~R\'akos$^{1,2}$ and M.~Paessens$^1$}

\address{$^1$ Institut f\"{u}r Festk\"{o}rperforschung, Forschungszentrum
J\"{u}lich, 52425 J\"{u}lich, Germany.}
\address{$^2$ Department of Physics of Complex Systems, Weizmann Institute of Science, Rehovot 76100, Israel}

\begin{abstract}
We investigate one-dimensional driven diffusive systems where particles may also be created and annihilated in the bulk with sufficiently small rate. 
In an open geometry, i.e., coupled to particle reservoirs at the two ends, these systems can exhibit ergodicity breaking in the thermodynamic limit. 
The triggering mechanism is the random motion of a shock in an effective potential. Based on this physical picture we provide a simple condition for the existence of a non-ergodic phase in the phase diagram of such systems. 
In the thermodynamic limit this phase exhibits two or more stationary states. However, for finite systems transitions between these states are possible. 
It is shown that the mean lifetime of such a metastable state is exponentially large in system-size. 
As an example the ASEP with the $A\varnothing A\rightleftharpoons AAA$ reaction kinetics is analyzed in detail. 
We present a detailed discussion of the phase diagram of this particular model which indeed exhibits a phase with broken ergodicity. 
We measure the lifetime of the metastable states with a Monte Carlo simulation in order to confirm our analytical findings.
\end{abstract}


\section{Introduction}
The closely related questions of phase coexistence and ergodicity
breaking in noisy one-dimensional systems are intriguing and have received wide
attention in the context of driven diffusive systems
\cite{Evan95,Evan98,Arnd98,Brau98,Barl98,Kafr02,Helb99,Schu01,Levi04}.

One-dimensional systems with short-range interaction and finite local state space do not show phase transition in thermal equilibrium. However, systems far from equilibrium may exhibit phase coexistence as well as ergodicity breaking \cite{Evan95,Evan98,Arnd98,Kafr02,Helb99}. 

In \cite{Parm03} a new type of phase coexistence was shown to occur in a driven diffusive system with weak non-conservation which has no analogue in conserving systems. Subsequently it was demonstrated \cite{Rako03} that similar systems may show diverse behavior: namely, a slight modification in the transition rates can lead to the appearance of a novel phase in which ergodicity is broken. In this paper we give a detailed description of a broad class of models including that of \cite{Parm03} and \cite{Rako03} and present a simple criterion which can be used to determine whether a given model in this class (described later) shows ergodicity breaking or not. As an example we study analytically and numerically the model of \cite{Rako03} in detail. 

The particular model considered in \cite{Parm03}  is the Totally Asymmetric Simple Exclusion Process (TASEP) with open boundaries augmented with a creation-annihilation dynamics on each lattice site. 
The TASEP is a well-known and probably the simplest nontrivial model of particles with biased diffusion on a one-dimensional lattice. Each lattice site can be either occupied ($A$) or empty ($\varnothing$), double occupancy is not allowed. Particles can only hop to the right\footnote{in a more general version (ASEP) particles can also hop in the opposite direction with smaller rate} with a given rate provided that the target site is empty. Particles are injected at the first and removed at the last site at given rates.

In addition to the biased diffusion of particles on an open lattice the model of \cite{Parm03} includes a simple creation-annihilation dynamics in the bulk
\begin{equation}
\label{creation-annihilation}
\0 \rightleftharpoons A
\end{equation} 
with rates $\omega_a$ (creation) and $\omega_d$ (annihilation) on each lattice-site. The mechanism is inspired by motor proteins moving along actin filaments, where attachment and detachment is possible. Earlier this model was introduced as a toy model reproducing stylized facts in limit order markets \cite{Will02}.

The phase diagram of the above system exhibits a phase where a domain of low particle density coexists with a domain of high density. The domain wall, i.e. a sharp interface between a high and a low density region, is localized in the bulk \cite{Parm03,Popk03,Evan03,Juha04}. In this phase the two different domains do not represent two possible {\it  global} steady states. The process is ergodic even in the thermodynamic limit and no hysteresis is possible. The existence of such a phase was very recently confirmed experimentally in a system of single-headed molecular motors (KIF1A) moving along a microtubule \cite{Nish05}.

In the following we show that in similar models, i.e. models with biased diffusion plus non-conserving reaction-kinetics, an additional phase (or phases) can be built up in which ergodicity is broken \cite{Rako03}. Here the high- and low-density phases do not coexist (like in \cite{Parm03}): either the one  or the other is present. For finite systems transitions between these states are possible but the transition times are exponentially large in system-size. The main dynamical mode, which determines the behaviour, is the random walk of a shock in an effective potential \cite{Rako03}. This one-particle collective mode provides a unified framework for understanding phase coexistence and ergodicity breaking in a certain class of systems. Based on this physical picture one can deduce the phase diagram of the particular model under consideration.

Ergodicity breaking is often associated with spontaneous symmetry breaking (SSB). However, the former is more general and one can speak about SSB only if the two (or more) stationary states are related by some symmetry. In the class of systems we consider this is usually not the case (another example for ergodicity breaking but not SSB is the Ising model with nonzero external field below the critical temperature).

The article is organized as follows: In section \ref{sec:hydro} we give some introduction to the hydrodynamic description we use for a class of reaction-diffusion systems. In section \ref{sec:shock} we present the effective dynamics, which is a simple random walk of the shock position in an energy-landscape. This physical picture leads to a criterion for the existence of a non-ergodic phase. A particular model, which satisfies this criterion is analyzed in detail in section \ref{sec:model}. Here we also compare the analytical predictions to Monte Carlo results. Finally, conclusions are drawn in section \ref{sec:conclusions}.

\section{Hydrodynamic description}
\label{sec:hydro}

In the following we restrict ourselves to one-species lattice gas systems in one dimension, but we note that considering models with continuous space would not make a big change. The description we present here is based on a hydrodynamic approach within Eulerian scaling. In this scale the lattice spacing $a\to 0$ leading to a continuous space-variable $x=ka$ and simultaneously the time is also rescaled as $t=a t_\mathrm{microscopic}$. Conserving driven particle systems on Euler-scale can usually be exactly described by partial differential equations for the local particle density \cite{Spoh91}:
\begin{equation}
\label{hidro}
\frac{\partial \rho(x,t)}{\partial t}+\frac{\partial j(\rho(x,t))}{\partial x}=0,
\end{equation}
where $j(\rho)$ is the macroscopic current which one would observe in an infinite system (or on a ring) in the stationary state with uniform particle density $\rho$. 
In a continuum description of interacting particle systems normally there is a term (referred to as diffusive or viscosity term) including a second derivative of the density. Note that on Euler scale this term vanishes since it has a prefactor proportional to the lattice spacing $a$.

It is usually a highly non-trivial question how to determine $j(\rho)$ for a given model and we have to emphasize that the mean-field approximation usually gives a wrong answer. In most cases neglecting short range correlations would not reproduce the correct $j(\rho)$ even qualitatively since these correlations play the major role in this question. However, there are systems where the homogeneous product measure (i.e. a factorized state without correlations) is stationary: in this case the mean-field, by coincidence, would give the right answer but this is rather the exception than the rule (the ASEP falls into this class). As an analogy we mention that the first order Taylor expansion also turns out to be exact when ``approximating'' linear functions. In this analogy improved mean-fields are similar to higher order Taylor expansions. 

Unfortunately there is a wide-spread misunderstanding about equation (\ref{hidro}) according to which this is often referred to as a kind of ``continuous mean-field'' description. The fact that there is a closed equation for the density does not mean that correlations are neglected, as one might think, but rather they are built in $j(\rho)$ which is indeed nothing else but a certain correlation function. Another common objection is that noise is neglected. Of course, noise is present on microscopic scale but this is scaled away on Euler-scale. To summarize: if $j(\rho)$ is exact then (\ref{hidro}) is also exact within Eulerian scaling (and has nothing to do with any kind of mean-field).

The systems we consider in this article can be constructed in general as follows: 
\begin{enumerate}
\item \label{point1} Take a one-species driven diffusive system in one dimension (which can be described by (\ref{hidro})) and couple it to particle reservoirs at the two ends. 
\item \label{point2}Introduce non-conserving dynamics in the bulk with reaction rates proportional to the inverse system-size. This rescaling of reaction rates is essential to obtain nontrivial new behaviour. If these rates are of larger order than $1/L$ then the steady state of the system is governed only by these reactions in the case of $L\to\infty$ and the boundaries do not play any important role. In the opposite case, when they are smaller than ${\cal O}(1/L)$, the system behaves like the original model without bulk non-conservation \cite{Parm03,Popk03,Evan03} (except on the transition line \cite{Juha04}). 
\end{enumerate}
As it was argued in \cite{Popk03} these systems on the Euler-scale (which now include the rescaling of bulk reaction rates according to point (\ref{point2})) can be described by the hydrodynamic equation 
\begin{equation}
\label{hidro2}
\frac{\partial \rho(x,t)}{\partial t}+\frac{\partial j(\rho(x,t))}{\partial x}=S(\rho(x,t)),
\end{equation}
where $S(\rho)$ is some kind of source term coming from the reaction kinetics in the bulk. The assumption this description is based on is that in the limit where $L\to\infty$ the probability of a non-conserving process to happen on a finite segment of the chain goes to zero. Consequently the system has enough time to relax (locally) to the stationary structure of the state which one would obtain with homogeneous density $\rho$ and without non-conserving rates, meaning that the local correlations are the same as those in the corresponding conserving system. This implies that the $j(\rho)$ appearing in (\ref{hidro}) is not effected by the bulk reaction-kinetics. 

For the correct evaluation of $S(\rho)$ it is also important to know the local correlations. Suppose that in addition to a given driven diffusive particle-conserving dynamics (point (\ref{point1}) above) the non-conserving processes from the local configuration $\cal C$ to $\cal C'$ with rates $r(\cal C',\cal C)$ are included (point (\ref{point2})). In the hydrodynamic limit we keep 
\begin{equation}
\label{rate_rescaling}
R({\cal C', C})=r({\cal C', C})L
\end{equation}
 fixed while the system-size $L$ is increased. Let $P({\cal C}, \rho)$ denote the probability of finding locally a specific configuration $\cal C$ in the stationary state of the conserving system with homogeneous density $\rho$. Knowing $P({\cal C}, \rho)$ (which requires the knowledge of local correlations) one can easily obtain $S(\rho)$ as follows:
\begin{equation}
\label{Source}
S(\rho)=\sum_{\cal C, C'} k({\cal C', C})R({\cal C', C})P({\cal C}, \rho),
\end{equation}
where $k({\cal C', C})$ simply denotes the particle number difference between configurations $\cal C$ and $\cal C'$.

As an example we give these quantities for the TASEP equipped with the simple creation-annihilation dynamics (\ref{creation-annihilation}) which was considered in \cite{Will02, Parm03, Evan03}. The TASEP falls into the class of models having product measure stationary states (i.e., without correlations) on an infinite chain, which implies $j(\rho)=\rho(1-\rho)$. On the other hand, with the notation $\Omega_a=R(A,\varnothing)$ and $\Omega_d=R(\varnothing,A)$, the source term is $S(\rho)=\Omega_a(1-\rho)-\Omega_d\rho$, since $P(A,\rho)=1-P(\varnothing,\rho)=\rho$. We  note that (a) including partial asymmetry in the hopping dynamics would only introduce a prefactor in the current (b) Due to the reasons described above in this specific case a mean-field ansatz would also lead to equation (\ref{hidro2}) with the same $j(\rho)$ and $S(\rho)$ \cite{Parm03}.

\subsection*{Stationary solution}
By requiring time-independent solutions, equation (\ref{hidro2}) reduces to the ordinary first order differential equation
\begin{equation}
\label{eq_DE}
\frac{d}{dx}j(\rho(x))\equiv v_c(\rho(x))\rho'(x)=S(\rho(x)),
\end{equation}
where $v_c(\rho)=\frac{d j(\rho)}{d \rho}$ is the collective velocity.
This equation allows for the calculation of the steady state density profile $\rho(x)$.  As boundary conditions the two reservoirs fix the densities at $x=0$ and $x=1$ (in the hydrodynamic limit we choose $a=1/L$ so the continuous space variable $x$ runs from 0 to 1) but as the differential equation (\ref{eq_DE}) is only of first order the solution usually fits at most one of the boundaries.  This discrepancy is resolved by the appearance of discontinuities which can be located at the boundaries ($x=0,1$) or in the bulk. Usually the way of solving (\ref{eq_DE}) numerically with given boundary conditions is the following. First one has to regularize the equation introducing a viscosity term proportional to $\epsilon$ which contains a second order derivative. This allows one to solve the resulting equation (since it became second order) and fit both boundaries. Finally one lets $\epsilon\to 0$ while usually one or more discontinuities appear in the solution. Nevertheless, the resulting curve is a solution of (\ref{eq_DE}) apart from these discontinuities.

However, in practice one does not have to perform the above program for all the possible choices of boundary conditions. In \cite{Popk03} some rules are given to decide which are the stable discontinuities at the boundaries. These are the consequences of the assumption that the hydrodynamic description is based on, and for conserving systems they reduce to the current minimization/maximization principle derived in \cite{Popk99, Hage01}. 

On the other hand, to have a localized shock in the bulk at $x_0$ it is necessary to satisfy
\begin{equation}
\label{stat_cond}
j\left(\rho(x_0-0)\right)=j\left(\rho(x_0+0)\right),
\end{equation}
which follows simply from mass conservation. Additionally, the structure of this shock has to be microscopically stable (e.g. a ``downward shock'' in the ASEP is not stable. For details see \cite{Hage01})

Given these rules it is still nontrivial to deduce the stationary density profile in general and the question remains whether there are cases when there is no unique solution, i.e. there are several possible stationary profiles which would imply ergodicity breaking. In the following we address this question by studying the dynamics of shocks in these systems.

\section{Shock dynamics -- Criterion for ergodicity breaking}
\label{sec:shock}

\subsection{Site dependent hopping rates} 
Suppose that there is a shock in the system which connects two solutions of (\ref{hidro2}): $\rho_\mathrm{left}(x)$ and $\rho_\mathrm{right}(x)$.  Even if it satisfies (\ref{stat_cond}) and it is a ``true'' localized shock (note that (\ref{stat_cond}) is only a necessary condition) its position fluctuates in time on the lattice scale. According to \cite{Kolo98} we assume that the shock position performs a simple random walk with effective hopping rates $D_\mathrm{left}$ and $D_\mathrm{right}$ which characterize the hopping to the left and right by one lattice site and are functions of the densities $\rho_\mathrm{right}$ and $\rho_\mathrm{left}$. In some special cases this effective one-particle description is exact \cite{Beli02,Kreb03,Bala04}. The average shock velocity 
\begin{equation}
\label{vs}
v_s=D_\mathrm{right}-D_\mathrm{left}
\end{equation}
 is easy to obtain from mass conservation (the non-conserving processes have different time-scale due to the rescaling of rates (\ref{rate_rescaling}) so they do not enter here):
\begin{equation}\label{vsx}
v_s(x)=\frac{j(\rho_\mathrm{right}(x))-j(\rho_\mathrm{left}(x))}{\rho_\mathrm{right}(x)-\rho_\mathrm{left}(x)},
\end{equation}
and the additional knowledge about the diffusion coefficient $D=(D_\mathrm{right}+D_\mathrm{left})/2$ uniquely defines the effective hopping rates. These rates for the ASEP are \cite{Kolo98}
\begin{eqnarray}
\label{hopping_rates}
D_\mathrm{left}(x)=\frac{j(\rho_\mathrm{left}(x))}{\rho_\mathrm{right}(x)-\rho_\mathrm{left}(x)},
\cr
D_\mathrm{right}(x)=\frac{j(\rho_\mathrm{right}(x))}{\rho_\mathrm{right}(x)-\rho_\mathrm{left}(x)}.
\end{eqnarray}
In a conserving system where the stationary solutions of (\ref{hidro}) are $\rho(x)=$const.\ the hopping rates are also constant all over the system. But as an effect of the nonconservative dynamics a source term $S(\rho)$ appears in (\ref{hidro2}) and leads to non-constant stationary solutions, therefore the hopping rates for the shock position also become space-dependent. Similar site dependent rates were used in \cite{Evan03}. 

Suppose that there is a point $x_0$ where $v_s(x_0)=0$, i.e., 
\begin{equation}
\label{zerov}
j(\rho_\mathrm{left}(x_0))=j(\rho_\mathrm{right}(x_0)). 
\end{equation}
There are two possibilities: the shock position at $x_0$ can be either stable or unstable. 
\begin{enumerate}
\item \label{casei}
Having a stable shock position requires $\frac{d}{dx}v_s(x)|_{x=x_0}<0$. In this case the shock position always has a drift towards the stationary point $x_0$. Differentiating the rhs.\ of (\ref{vsx}) and using (\ref{eq_DE}) and (\ref{zerov}) the above condition yields 
\begin{equation}
\frac{S(\rho_\mathrm{right}(x_0))-S(\rho_\mathrm{left}(x_0))}{\rho_\mathrm{right}(x_0)-\rho_\mathrm{left}(x_0)}
<0.
\end{equation}
This is equivalent to requiring
\begin{equation}
S(\rho_a)>S(\rho_b),
\end{equation}
where $\rho_a=\min(\rho_\mathrm{right}(x_0),\rho_\mathrm{left}(x_0))$ and $\rho_b=\max(\rho_\mathrm{right}(x_0),\rho_\mathrm{left}(x_0))$. 
\item \label{caseii}
On the other hand 
\begin{equation}
\label{usp}
S(\rho_a)<S(\rho_b)
\end{equation}
implies an unstable shock position. In this case small deviations from $x_0$ are amplified.
\end{enumerate}
In the following we show that unstable (fixed) shock positions lead to ergodicity breaking in the thermodynamic limit. 

\subsection{Effective potential}
\label{sec:effective_potential}

The simple random walk (i.e. only nearest neighbour jumps are allowed) of a single particle, i.e., the shock position in a confined volume in one dimension necessarily results in an equilibrium dynamics which can be characterised by an effective potential. This potential on the lattice scale satisfies
\begin{equation}
\label{discrete_E}
\exp(E_k-E_{k+1}) = \frac{p_{k}}{q_{k+1}},
\end{equation}
where $p_{k}$ is the hopping rate from site $k$ to $k+1$, while $q_{k}$ is the one from site $k$ to $k-1$. It can be shown that the potential in the hydrodynamic limit using the rates $D_\mathrm{left}(x)$ and $D_\mathrm{right}(x)$ becomes
\begin{equation}
\label{potential}
E(x)=L \int^x \ln\left(\frac{D_\mathrm{left}(x')}{D_\mathrm{right}(x')}\right) dx'.
\end{equation}
In the case of a stable shock position (case (\ref{casei})) $E(x)$ has a local minimum at $x_0$. The probability density to find the shock at position $x$ is proportional to $\exp(-E(x))$ which can be approximated by a Gaussian around $x_0$. The width of the distribution is proportional to $L^{1/2}$  on the lattice scale (and to $L^{-1/2}$ on Euler-scale) \cite{Evan03}.

In the opposite case (case (\ref{caseii})), when the shock position is unstable, the potential $E(x)$ has a local maximum at $x_0$ resulting in (at least) two local minima. Since $E(x)$ is proportional to $L$ the potential barrier between these metastable positions is proportional to $L$ too. From this one expects transition times which are exponentially large in $L$ leading to ergodicity breaking in the thermodynamic limit. In section \ref{sec:transitions} it is shown that for systems having two metastable shock positions at the two boundaries the average transition time between these states grows exponentially with the system size (with an algebraic prefactor). It is plausible to assume that the same is true when the metastable states are in the bulk. 

\subsection{Criterion for ergodicity breaking}

To conclude, the conditions which has to be satisfied for non-ergodic behavior are summarized below:

If one can find $\rho_a$ and $\rho_b$ such that 
\begin{enumerate}
\item \label{egy} $\rho_a < \rho_b$
\item \label{ketto} the shock connecting $\rho_a$ and $\rho_b$ (either $\rho_a|\rho_b$ or $\rho_b|\rho_a$) is stable
\item \label{harom} according to (\ref{zerov}) $j(\rho_a)=j(\rho_b)$ and
\item \label{negy} $S(\rho_a)<S(\rho_b)$,
\end{enumerate} 
then non-ergodic behavior is possible. Otherwise this kind of mechanism does not give rise to ergodicity breaking.

\subsection{Remarks}

To check whether the above criterion is met the specific form of the current density relation of the driven diffusive system has to be known. In this subsection we give some examples where conditions \ref{egy}-\ref{negy} can or cannot be satisfied.

1. Assume that the conserving part of the dynamics is the ASEP. The current is $j(\rho)\sim\rho(1-\rho)$ therefore according to condition \ref{harom}.\ $\rho_b=1-\rho_a$. The ASEP has a product measure stationary state on an infinite lattice (or with periodic boundaries). As it is argued above, weak non-conservation with rates of order $1/L$ (and the open geometry) does not change this property in the bulk in the $L\to\infty$ limit. Because of this factorisation every kind of reaction kinetics give a polynomial contribution to $S(\rho)$ the order of which is the number of lattice sites involved in the reaction. It is easy to show that with the physical constraint $S(0)\geq 0$ and $S(1)\leq 0$ first and second order polynomials cannot satisfy condition \ref{negy}. together with \ref{egy}. One needs to include at least a third order source term, and correspondingly a three-site reaction, for having ergodicity breaking (in the ASEP condition \ref{ketto}. is always satisfied).

2. It is also easy to see that every driven diffusive system equipped with a simple creation-annihilation dynamics (\ref{creation-annihilation}) is ergodic. The reason is that (\ref{creation-annihilation}) always gives a linearly decreasing $S(\rho)$ which cannot satisfy condition \ref{negy}. 

3. In general, however, it is not necessary to include three-site reactions to get ergodicity breaking. As an example it can be shown that in the $k$-step exclusion process \cite{Guio99} with the two-site reaction $\varnothing\varnothing \rightleftharpoons A\varnothing$, the above conditions can be satisfied. In this system the steady state is factorized, like in the ASEP, but the current is not a symmetric function of the density, therefore the jumps do not have to be symmetric with respect to $\rho=1/2$.

In the following, because of its simplicity, we study in detail the ASEP with the reaction $A\varnothing A\rightleftharpoons AAA$.

\section{An example: the model of ref.\ \cite{Rako03}}
\label{sec:model}

The model we consider in this section was introduced in \cite{Rako03}. It was shown that the phase diagram exhibits a phase where ergodicity is broken and the underlying mechanism, i.e., the random motion of a shock in an effective potential was also presented. Here we present a more detailed discussion of the phase diagram of this model and measure the lifetime of the metastable states in the non-ergodic phase to be compared to the analytical prediction.

\begin{figure}
\vspace{3mm}
\centerline{\epsfig{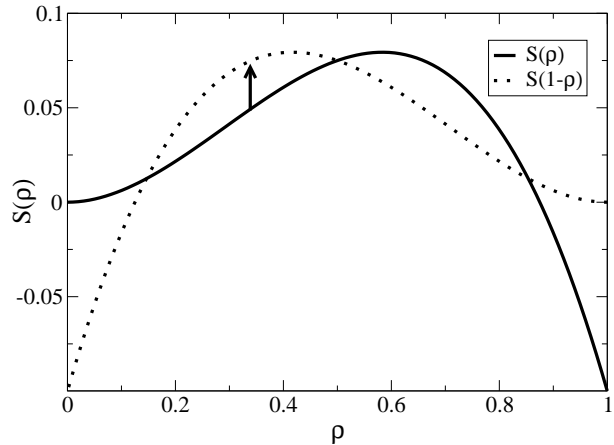}}
\caption{The source term $S(\rho)$ compared to $S(1-\rho)$ for $\Omega_a=0.7$ and
$\Omega_d=0.2$: There exists a $\rho_a<1/2$ such that
$S(\rho_a)<S(1-\rho_a)$. 
}\label{fig2}
\end{figure}

In the model we consider in the following the driven diffusive dynamics is defined by the TASEP, but including backwards hoppings with smaller rate (ASEP) would not change the behaviour. The current which enters the hydrodynamic equation is 
\begin{equation}
j(\rho)=\rho(1-\rho),
\end{equation}
which gives the collective velocity
\begin{equation}
v_c(\rho)=1-2\rho.
\end{equation}
As argued above at least three point reactions have to be coupled to the ASEP to allow for unstable shock positions. In what follows we consider the reaction kinetics \cite{Rako03}
\begin{equation}
A\0 A\ \rightleftharpoons AAA,
\label{eq_react}
\end{equation}
with rates $\omega_a$ and $\omega_d$. This may be interpreted as activated Langmuir kinetics, and yields a  source term
\begin{equation}
S(\rho)=\Omega_a \rho^2 (1-\rho)-\Omega_d \rho^3,
\end{equation}
where $\Omega_{a,d}=\omega_{a,d}L$ are the rescaled rates according to (\ref{rate_rescaling}). As illustrated in figure \ref{fig2} for certain choice of rates this source term fulfills the conditions for the existence of unstable shock positions. At the first lattice site particles are created at rate $\rho_-$ when this site is empty, while at the last site they are annihilated at rate $1-\rho_+$. These boundary rates represent particle reservoirs at the two boundaries with densities $\rho_-$ and $\rho_+$.

\begin{figure}
\vspace{3mm}
\centerline{\epsfig{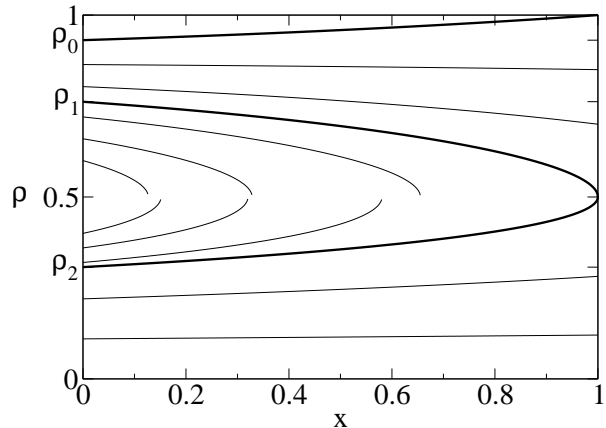}}
\caption{The flow-field $\rho(x)$ of the differential equation (\ref{ode}). 
The parameters are $\Omega_a=0.7$ and $\Omega_d=0.1$. 
The solutions are monotonically increasing and convex in the regions 
$0<\rho<1/2$ and $\rho^*<\rho\leq 1$ while they are decreasing and concave for 
$1/2<\rho<\rho^*$, where $\rho^*=\Omega_a/(\Omega_a+\Omega_d)=0.875$ is the stationary density of the reaction-kinetics (\ref{eq_react}) without biased diffusion. Since the collective velocity $v_c(\rho)$ is zero at $\rho=1/2$ 
the solutions are singular at this point.}
\label{fig3}
\end{figure}

In order to calculate the stationary density profiles we rewrite the hydrodynamic equation (\ref{eq_DE}) as
\begin{equation}
\label{ode}
\rho'(x) = \frac{S(\rho)}{v_c(\rho)}.
\end{equation}
Integration directly yields the implicit formula for the flow-field (the set of $\rho(x)$ curves satisfying the ordinary first order differential equation (\ref{ode})):
\begin{equation}
\label{solution}
x(\rho)=-\frac{1}{\Omega_a\rho}+\frac{\Omega_a-\Omega_d}{\Omega_a^2}
\ln\left|\frac{1}{\rho^*}-\frac{1}{\rho}\right|+c
\end{equation}
where $\rho^*= (\Omega_a+\Omega_d)/\Omega_a $ is the stationary density of the reaction kinetics (\ref{eq_react}) without biased diffusion and $c$ is an integration constant. The corresponding flow-field $\rho(x)$ is shown in figure \ref{fig3}. The direct calculation of the inverse function is possible in terms of the Lambert-functions but unfavorable. 

\subsection{Phase diagram}

\begin{figure}
  \centerline{\epsfig{file=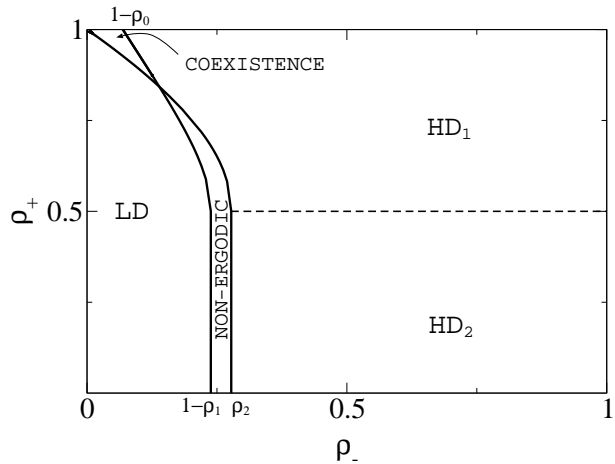,width=9truecm}}
  \caption{Phase diagram for fixed $\Omega_a=0.7$ and $\Omega_d=0.1$ with two high-density phases (HD1, HD2), a low-density phase (LD), a coexistence phase, and the non-ergodic phase.}
  \label{fig:phase_diagram}
\end{figure}

The solutions (\ref{solution}) enable us to deduce the steady-state phase diagram of the system in terms of the boundary densities $\rho_-$ and $\rho_+$ (see figure \ref{fig:phase_diagram}).

Before the discussion of the phase diagram let us define some special values of the left boundary density, which correspond to special lines of the flow-field. If the right boundary density is set to 1/2 there are two solutions which satisfy this boundary condition due to the singularity at $\rho=1/2$. We call $\rho_{1(2)}$ the value of the upper (lower) solution at $x=0$ (see figure \ref{fig3}). In addition we define $\rho_0$ as the starting point of the solution fitting the right boundary density $\rho_+=1$ (see figure \ref{fig3}).

\begin{figure}
  \centerline{\epsfig{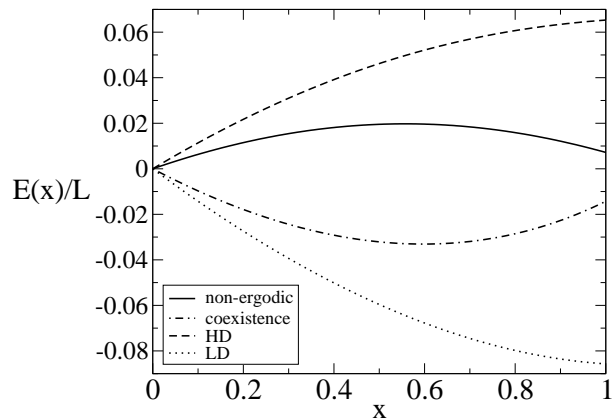}}
  \caption{Examples of the effective potential in four phases. Note that in the HD and LD $E(x)$ can be either convex or concave. Parameters are: $\Omega_a=0.7 \ \Omega_d=0.1$; $\rho_-=0.25 \ \rho_+=0.6$ (non-ergodic),$\rho_-=0.1 \ \rho_+=0.903$ (coexistence), $\rho_-=0.27 \ \rho_+=0.6$ (HD), $\rho_-=0.1 \ \rho_+=0.89$ (LD).}
  \label{fig:energy_profiles2}
\end{figure}

First we consider the case $\rho_-<\rho_2$ and $1/2<\rho_+$. In this case $\rho_\mathrm{left}(x)$ and $\rho_\mathrm{right}(x)$ are the solutions (\ref{solution}) with the boundary conditions $\rho_\mathrm{left}(0)=\rho_-$ and $\rho_\mathrm{right}(1)=\rho_+$ and they are well-defined in the whole range of $x$. As $\rho_\mathrm{right}(x)>1/2>\rho_\mathrm{left}(x)$ the shock connecting the two solutions is stable so one can define the effective potential $E(x)$ for the shock position according to (\ref{potential}). In figure \ref{fig:energy_profiles2} four qualitatively different energy-landscapes are shown which are present in the region of the phase diagram under consideration. They are calculated by inserting the analytical solutions of $\rho_\mathrm{left}(x)$ and $\rho_\mathrm{left}(x)$ in into (\ref{potential}). These effective potentials indicate four qualitatively different behaviours: 
\begin{itemize}
\item
There is a region in the parameter space where $E(x)$ is monotonically decreasing resulting in a stationary shock position at the right boundary. Consequently the bulk solution is given by $\rho_\mathrm{left}(x)$ -- this is the low density phase (LD): 
\item
In another region, called high density phase (HD), the potential $E(x)$ is monotonically increasing, therefore the stationary shock position is at the left boundary and the solution is $\rho_\mathrm{right}(x)$ in the bulk. 
\item 
In the coexistence phase $E(x)$ has a local minimum in the bulk leading to a stable shock position at a macroscopic distance from both boundaries. The shock separates a low density - and a high density region characterised by  $\rho_\mathrm{left}(x)$ and $\rho_\mathrm{right}(x)$. This phase is analogous to the one found in \cite{Parm03}. When measuring the density profile by a time average one observes a smooth curve connecting the two regions. This is the consequence of the fluctuation of the shock position, the shock itself is sharp. The width of this ``transition region'' can be calculated by a harmonic approximation of the potential in the vicinity of the minimum, which gives an $L^{1/2}$ dependence on the lattice scale (on the Euler-scale the width vanishes as $L^{-1/2}$). 
\item
In addition a ``non-ergodic'' phase is found, which is characterised by a potential with a local maximum in the bulk. The two minima at the left and right boundary correspond to the two meta-stable states: $\rho_\mathrm{left}(x)$ and $\rho_\mathrm{right}(x)$. For finite systems a transition between the low-density and the high-density state is possible but very improbable if $L$ is large due to the potential barrier $\Delta E\sim L$ (see section \ref{sec:effective_potential}). Of course strictly speaking for any finite $L$ there is an unique stationary state which is some kind of linear combination of the high - and low density states but the relaxation time to this stationary state diverges exponentially with $L$. In the limit $L\to \infty$ the system becomes non-ergodic since one of the stable states is selected in a very short while (depending mainly on the initial configuration) and a transition to the other one is not possible, therefore a time-average does not reproduce the ensemble-average. 
\end{itemize}

With the picture of the moving shock in mind it is also possible to derive exactly the two lines in the phase diagram shown in figure \ref{fig:phase_diagram} which separate these phases. One of them is determined by the condition $E'(0)=0$. To the right (left) of this line a shock at the left boundary -- and therefore the high density state -- is stable (unstable). Along the other line $E'(1)=0$, and it separates the two regions where the shock at the right boundary -- and therefore the low density state -- is stable (to the left) or unstable (to the right). The coexistence phase is the one where neither of them are stable and as a consequence the shock is driven into the bulk and is localised there. In the non-ergodic phase both the low density and the high density state are stable. 

The sign of the slope of the effective potential is determined by the sign of the average shock velocity $v_s(x)$ (see equation (\ref{vs},\ref{potential})) so in what follows we focus on this quantity at the two boundaries. The shock velocity at the left boundary $v_s(0)$ is zero if $j(\rho_\mathrm{left}(0))=j(\rho_\mathrm{right}(0))$ and therefore
\begin{equation}
\label{line1}
\rho_-\equiv \rho_\mathrm{left}(0)=1-\rho_\mathrm{right}(0),
\end{equation}
where $\rho_\mathrm{right}(x)$ is the solution fitting the right boundary density $\rho_+$. Increasing the value of $\rho_-$, $v_s(0)$ decreases so the high density phase is stable to the right of the line defined by (\ref{line1}). This phase boundary starts at $\rho_-=1-\rho_0, \rho_+=1$ and ends at $\rho_-=1-\rho_1, \rho_+=1/2$

The analogous equation for the phase boundary of the stable low-density state is
\begin{equation}
\rho_\mathrm{left}(1)=1-\rho_\mathrm{right}(1)\equiv 1-\rho_+,
\end{equation}
where $\rho_\mathrm{left}(x)$ is the solution fitting the left boundary density $\rho_-$. For similar reasons the low-density state is stable to the left of this line, which starts at $\rho_-=0, \rho_+=1$ and ends at $\rho_-=\rho_2, \rho_+=1/2$. Note that the condition for having a non-ergodic phase is $\rho_2>1-\rho_1$, which is not satisfied for all values of $\Omega_{a,d}$.

\begin{figure}
\centerline{\epsfig{file=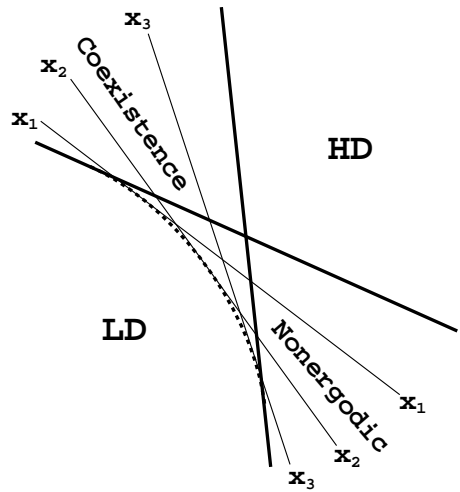,width=6truecm}}
\caption{Shown is the narrow phase near the intersection point of the two nontrivial curves of the phase diagram (figure \ref{fig:phase_diagram}). The two thick solid lines represent these phase boundaries in this schematic enlarged figure. Along the thin solid lines the energy profile has a stationary point at a given $x$ ($x_1>x_2>x_3$). In the coexistence phase this is a local minimum whereas in the non-ergodic phase this is a local maximum. The second derivative changes sign on the thick dashed curve. In the ``triangular'' area the energy landscape has a local minimum and a local maximum as well in the bulk.}\label{fig:narrow}
\end{figure}

\begin{figure}
  \centerline{\epsfig{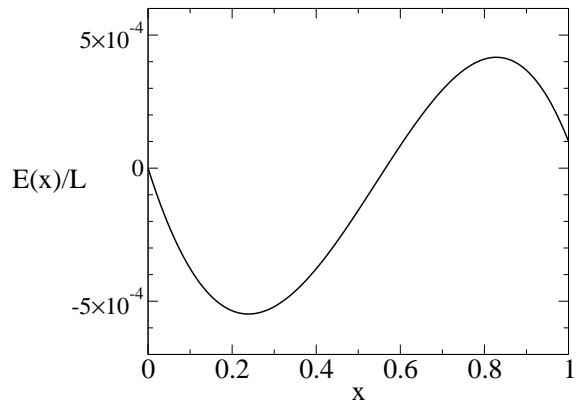}}
  \caption{The potential in the narrow phase near the intersection point of the lines separating the HD, LD, coexistence, and non-ergodic phases. Due to the small energy-barrier one would need huge systems to observe this phase by a Monte-Carlo simulation (compare the scale on the ``y'' axis to the one of figure \ref{fig:energy_profiles2}). Parameters are: $\Omega_a=0.7\ \Omega_d=0.1\ \rho_-=0.138 \ \rho_+=0.8435$}
  \label{fig:strangeE}
\end{figure}

We note that the analytical treatment suggests an additional phase in the vicinity of the intersection point of the two lines discussed above (see figure \ref{fig:narrow}). In this phase $E(x)$ has a local maximum and a local minimum as well (see figure \ref{fig:strangeE}). This type of energy-landscape also leads to ergodicity breaking. In one of the stationary states the shock is located in the bulk, in the other it is at the right boundary (LD). Although this phase is very narrow and not observable by Monte-Carlo simulations in our specific model, it indicates that in general $E(x)$ can be more complex and can have more stationary points. 

Now let us consider another region of the phase diagram. Suppose that keeping $\rho_+$ fixed between 1/2 and 1, we increase $\rho_-$ beyond $\rho_2$. Due to the singularity the left solution $\rho_\mathrm{left}(x)$, and therefore $E(x)$, is not well-defined in the whole range of $x$ anymore (see figure \ref{fig3}). Nevertheless it is perfectly defined in a macroscopic interval $x\in (0,z)$, where $0<z<1$ is the point where $\rho_\mathrm{left}(x)$ becomes singular. However, due to the negative average speed the shock stays in this region so there is no change in the qualitative behaviour. If $\rho_-$ approaches 1/2 the length of the interval where $E(x)$ is defined shrinks to zero. Further increasing of $\rho_-$ leads to the loss of the effective potential. Although the bulk behaviour does not change in this case we expect a change in the details in the way the density profile approaches the bulk solution on the lattice scale near the left boundary. 

The situation is similar to the case in the (conserving) ASEP in the high density phase. If the left boundary density $\rho_-$  is less then 1/2 there is a ``pure'' shock localized at the left boundary leading to a simple exponential decay to the bulk density, which is equal to the right boundary density $\rho_+$. However, if the left boundary density is above 1/2 there is an algebraic prefactor to the exponential \cite{Derr93a,Schu93}, which can be interpreted as a result of a combination of an algebraic decay to the maximal current state $\rho=1/2$ and a shock from this value to the bulk $\rho=\rho_+$ (for details see \cite{Kolo98}). Similarly, in our model we expect a pure exponential in the approach to the bulk $\rho(x)$ in the HD for $\rho_-<1/2$, while for $\rho_->1/2$ an algebraic modification is expected. 

Nevertheless the bulk behaviour, which is given by $\rho_\mathrm{right}(x)$, does not change in the whole region $\rho_+>1/2, \rho_->\rho_2$ so this belongs to the HD. 

The above situation is analogous to the one where being in the LD we decrease $\rho_+$ with fixed $\rho_-$. Although $E(x)$ could be defined also in the case of $\rho_+<1/2$ since both solutions $\rho_\mathrm{left}(x)$ and $\rho_\mathrm{right}(x)$ are regular, an algebraic modification of an exponential decay to the bulk profile is expected, just like in the usual ASEP in the LD if $\rho_+<1/2$ \cite{Derr93a,Schu93}. The bulk behaviour however, remains unchanged and is governed by $\rho_-$ in this region so this belongs to the LD.

Finally suppose that the system is in the HD discussed before and we decrease $\rho_+$. As $\rho_+$ approaches 1/2 the bulk profile converges to the solution which starts at $\rho_1$ and ends at 1/2. We argue that further decreasing of $\rho_+$ does not change the bulk behaviour. This case is analogous to what happens in the conserving ASEP in the HD-maximal current (MC) phase boundary. There, an algebraic decay to the bulk density 1/2 sets in if $\rho_+<1/2$. Here, we expect similar behaviour near the left boundary, but it is still not trivial to predict the type of decay because this algebraic decay is somehow combined with the singularity of the bulk profile at $x=1$. Consequently, in the whole range of parameters $\rho_+<1/2$ and $\rho_->\rho_2$ the bulk stationary profile is independent of both boundaries (and is given by the thick line starting at $\rho_1$ on figure \ref{fig3}). In this sense this is similar to the MC phase in the ASEP phase diagram and therefore we distinguish between the HD1 ($\rho_+>1/2$) and HD2 ($\rho_+<1/2$) phases.

The region $1-\rho_1<\rho_-<\rho_2$ and $\rho_+<1/2$ belongs to the non-ergodic phase. The stationary state here is either the corresponding low-density state determined by $\rho_-$ or the state corresponding to the HD2. Summarizing, we can say that the bulk behaviour below the $\rho_+=1/2$ line is just the same as on this line itself.

\subsection{Transition times} 
\label{sec:transitions}

The existence of two metastable states is not a sufficient condition for ergodicity breaking. Here we show that the transitions from one state to the other are impossible in the thermodynamic limit. This is done by showing that the mean first passage times (MFPT) of the shock from one boundary to the other grows exponentially with the system size (which is suggested by the linearly growing potential barrier).

The MFPT $\tau$, of a random walker with site-dependent hopping rates from site $1$ to $L$ can be calculated using a formula derived by Murthy and Kehr \cite{Murt89},
\begin{equation}
\label{mfpt}
\tau=\sum_{k=1}^{L-1} \frac{1}{p_k} + \sum_{k=1}^{L-2} \frac{1}{p_k} \sum_{i=k+1}^{L-1}
              \prod_{j=k+1}^{i}\frac{q_j}{p_j}
\end{equation} 
where $p_j$ is the hopping rate of the random walker from site $j$ to $j+1$ and $q_j$ from site $j$ to $j-1$. According to (\ref{discrete_E}) this can be written as
\begin{equation}
\label{mfpt2}
\tau=\sum_{1\leq k\leq l \leq L-1} \frac{e^{E_l-E_k}}{p_k}
\end{equation} 
 This formula reduces only in some special cases to a closed form and one has to rely on numerical evaluations in general.  We tested the scaling of $\tau$ with $L$ for several shapes of the energy landscape with a  maximum proportional to $L$. We found that the MFPT $\tau$ from site 1 to site $L$ is exponential in $L$ with an algebraic prefactor:
\begin{equation}
\label{Ediv}
\tau \sim L^\alpha \exp(\Delta E),
\end{equation}
where $\Delta E$ is the maximum of $E(x)-E(y)$ with the condition $x>y$, and is proportional to $L$. The value of $\alpha$ depends on the specific form of $E(x)$, more precisely on the behaviour near the minimum and maximum of it. In our case, when at the minimum (at the boundary) $E(x)$ starts linearly and at the maximum it is quadratic $\alpha=1/2$. 

The formula (\ref{Ediv}) is also supported by an analytical argument based on approximating the double sum of (\ref{mfpt2}) by a double integral:
\begin{equation}
\label{contlim}
\tau\approx L^2\int_{0<y<1} \int_{0<x<y}\frac{e^{L(\tilde E(x)-\tilde E(y))}}{D_\mathrm{right}(y)}dx dy.
\end{equation}
Here we introduced $\tilde E(x)=E(x)/L$, which remains constant with increasing $L$ and we used the right hopping rate $D_\mathrm{right}$ of the shock. These are functions of the continuous space-variable $x$. For energy profiles similar to the solid line of figure \ref{fig:energy_profiles2} the main contribution to (\ref{contlim}) comes from the part where $y$ is close to zero and $x$ is in the vicinity of the local maximum of $E(x)$. Using the first and second order Taylor-expansions of $\tilde E(x)$ near zero and $x_0$ (the point where $E(x)$ has its local maximum) and extending the limits of the integrals one gets
\begin{equation}
\tau\approx\int_0^\infty \frac{e^{-L(\tilde E(0)+y\tilde E'(0))}}{D_\mathrm{right}(y)}dy 
\int_{-\infty}^\infty e^{L(E(x_0)+\frac{1}{2}(x-x_0)^2\tilde E''(x_0))}dx.
\end{equation}
For large $L$ the $y$-dependence of $D_\mathrm{right}$ can be neglected leading to
\begin{equation}
\label{tauanal}
\tau\approx\frac{\sqrt{2\pi}}{\tilde E'(0)D_\mathrm{right}(0)\sqrt{\tilde E''(x_0)}} L^{1/2} e^{L\Delta \tilde E}.
\end{equation}
This is indeed of the form of (\ref{Ediv}) with $\alpha=1/2$. We note that for energy profiles which have different behaviour near the minimum and maximum, the same argument would typically lead to the same form with different value of $\alpha$. 

\begin{figure}
\vspace{3mm}
\centerline{\epsfig{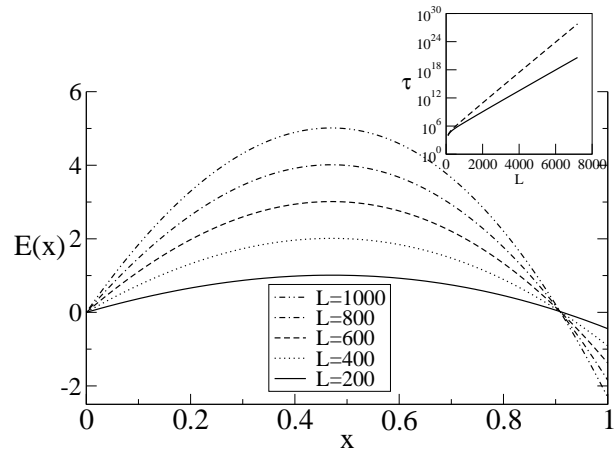}}
\caption{Energy profiles for $\rho_-=0.2705,\, \rho_+=0.63,\, \Omega_a=0.5,\, \Omega_d=0.1$ and different system sizes. The inset shows the mean first passage times from left to right (solid line) and from right to left (dashed line) against the system size.  }\label{fig4}
\end{figure}

\begin{figure}
\vspace{3mm}
\centerline{\epsfig{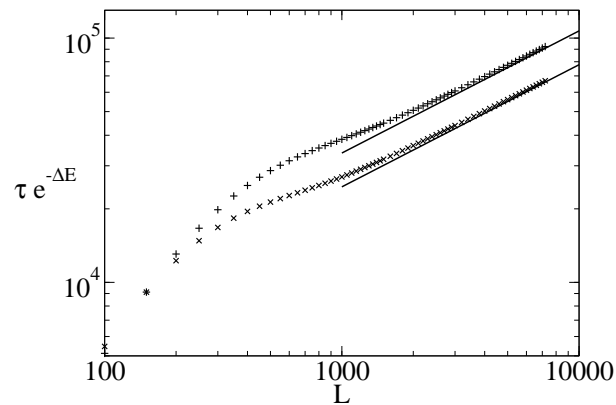}}
\caption{$\tau/\exp(\Delta E)$ is shown as a function of $L$ for the LD ($\times$) and HD ($+$) states. The parameters are: $\Omega_a=0.5, \Omega_d=0.1, \rho_-=0.2705, \rho_+=0.63$. The straight lines have slope 1/2 and are fitted by hand (note the logarithmic scale). This confirms the $L^{1/2}$ prefactor in $\tau$, which corresponds to $\alpha=1/2$.}\label{fig:tauperexpDE}
\end{figure}

The MFPTs can be calculated also numerically, using the hopping rates (\ref{hopping_rates}) which result from the density profiles suggested by the hydrodynamic description. The energy profiles and the resulting MFPTs for different system sizes are shown in figure \ref{fig4}. First the hydrodynamic equation (\ref{ode}) was solved numerically with the two different boundary conditions: $\rho(0)=\rho_-$ and $\rho(1)=\rho_+$ which gave the two solutions $\rho_\mathrm{left}(x)$ and $\rho_\mathrm{right}(x)$. After discretisation of $x$ the left and right hopping rates were defined by (\ref{hopping_rates}) which were used to calculate the energy profile (\ref{discrete_E}) and $\tau$ (\ref{mfpt}). It can be seen that the height of the energy barrier increases linearly with system size and this results in an exponential increase of the first passage times. It is also possible to check the (less important) algebraic prefactor. In figure \ref{fig:tauperexpDE} $\tau \exp(-\Delta E)$ is shown against $L$ and the prefactor $L^{1/2}$ is confirmed. However, the constant factor in (\ref{tauanal}) does not agree with the numerical results. It can often happen when a sum is replaced by an integral that the functional dependence of a parameter is reproduced correctly but not the constant factor. Figures \ref{fig4} and \ref{fig:tauperexpDE} confirm that the integral approximation (\ref{contlim}), resulting in (\ref{tauanal}), correctly describes the behavior of (\ref{mfpt}) for $L\to\infty$ up to a constant factor. A comparison of this predicted mean first passage time and simulation results of the full system is made in section \ref{sec:simresults}. 

We note that of course the potential $E(x)$ characterizes only the equilibrium behaviour of the random walker: it gives the probability distribution of the shock position along $x$ after relaxation, so for times $t\gg\tau$. In contrast, the MFPT is a quantity which refers to non-equilibrium and is not given by $E(x)$ only. Note that $\tau$ is a function of the hopping rates which are not uniquely defined by $E(x)$. However, it turns out that the behaviour of $\tau$ for large $L$ (\ref{Ediv}) does not depend strongly on the specific choice of the hopping rates, but it changes only the prefactor (gives a different time-scale).

To summarize, we conclude that the transition times of the shock from one boundary to the other diverge exponentially with system size and therefore ergodicity is broken in the thermodynamic limit.

\subsection{Simulation technique}
In order to confirm our predictions we performed Monte-Carlo (MC) simulations. In the effective dynamics described above the only degree of freedom is the position of the shock, which has to be defined on the lattice scale to be able to compare the analytical results with simulations. 

In the ASEP this is usually done by introducing second class particles ($B$) \cite{Ferr91}. These particles behave as normal ($A$) particles with respect to holes but they count as holes in an environment of $A$ particles. If there is only a single $B$ particle in the system the position of it is a good definition of the shock position on the lattice scale. In non-conserving systems, however, one has to take care of not losing the second-class particle (and not introducing an additional one). To this end we introduce another method to trace the shock position on the microscopic scale.  

\begin{figure}
  \centerline{\epsfig{file=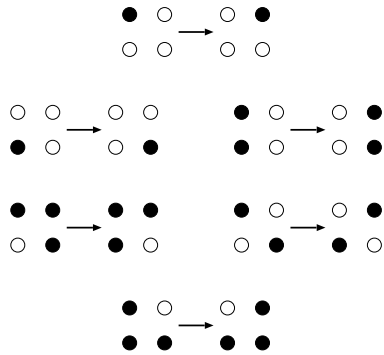,width=5truecm}}
  \caption{Hopping processes in the bulk. All
    of the above processes have rate one. On the upper lattice
    the dynamics of the original model is recovered while a single
    particle moving on a separate (lower) lattice was introduced to
    indicate the position of the shock. The reactions  
    ($\bullet\circ\bullet\rightleftharpoons\bullet\bullet\bullet$) take
    place  only
    on the upper lattice independently of the configuration of the
    lower one. Note that in our simulations the last hopping process does not
    happen since we introduce only one particle on the lower lattice.}
  \label{fig:hoppings}
\end{figure}

Consider an additional one dimensional lattice with $N$ sites parallel to the original (upper) one. The hopping processes are those explained in figure \ref{fig:hoppings} while the bulk non-conserving part of the dynamics ($\bullet\circ\bullet\rightleftharpoons\bullet\bullet\bullet$) as well as the injection and ejection at the boundaries act only on the upper lattice. One can see that the dynamics of the original model is recovered on the upper lattice.  

Notice that without non-conserving processes in the bulk and identifying 
\begin{equation}
  \label{map}
  \begin{array}{c} \bullet \\ \circ \end{array} \to A, \qquad
  \begin{array}{c} \circ \\ \circ \end{array} \to \varnothing, \qquad
  \begin{array}{c} \bullet \\ \bullet \end{array} \to B, \qquad
  \begin{array}{c} \circ \\ \bullet \end{array} \to B
\end{equation}
one recovers the ASEP with second-class particles \footnote{Alternatively one can consider the configuration ${\circ \choose \bullet}$ as a distinct ($C$) type of particle. Since this counts as an empty site if it meets $A$ or $B$ but behaves as a particle in an empty environment $C$ could be called a third-class particle.}. The above postulated dynamics also serve as an alternative definition of the second class particle in an open system (there is a minor difference between these dynamical rules and those of \cite{Kreb03} at the boundaries). 

Introducing only one single particle on the lower lattice makes it possible to trace the shock position without having any effect on the (non-conserving) dynamics of the original system. Furthermore the setup described above has another practical advantage, namely, it is suitable to adapt multispin coding \cite{Bark99}, which speeds up the simulations remarkably.

\subsection{Monte Carlo Results}
\label{sec:simresults}
\begin{figure}
\centerline{\epsfig{file=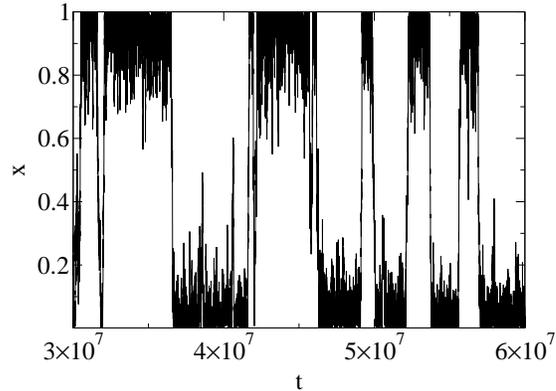,width=6truecm,angle=270}}
\caption{Snapshot of the time evolution of the scaled position of the
  second class particle for $L=1000, \rho_{-}=0.2705, 
\rho_{+}=0.63, \Omega_a=0.5, \Omega_d=0.1$. A position of the second
class particle near the
left boundary ($x\approx 0$) corresponds to the high density state,
while a position near the
right boundary ($x\approx 1$) corresponds to the low density state.
}
\label{fig:timewindow}
\end{figure}

In the simulations we measured the time dependence of the position of the shock (the particle on the lower lattice) in the broken ergodicity phase. A time window of the result is shown in figure \ref{fig:timewindow}. One can see that the shock spends most of the time in the vicinity of either boundary: the shock being at the left/right boundary means that the system is in the HD/LD state. 

\begin{figure}
\centerline{\epsfig{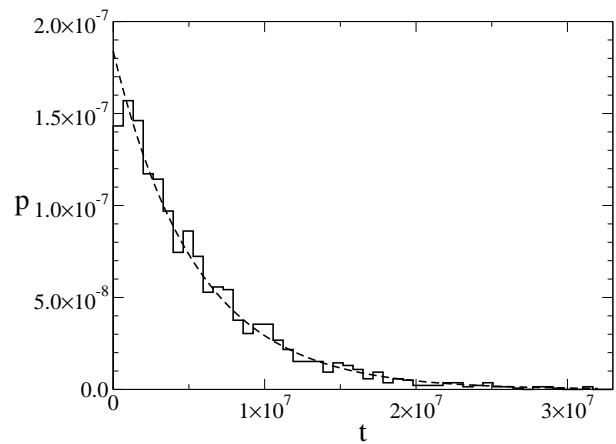}}
\caption{Numerically determined distribution of the transition times from the high density to the low density state (histogram) compared to the exponential distribution (dashed line). Similar results are found for the transition times in the other direction but with a different mean. Parameters are: $L=1000, \rho_-=0.2705, \rho_+=0.63, \Omega_a=0.5, \Omega_d=0.1$. The histogram is based on a list of 2094 measured transition times.}
\label{fig:distribution}
\end{figure}

We also measured the first passage times of the shock from left to right and vice versa which correspond to the lifetime of the HD and LD states. The one-particle picture implies that the first passage times have to follow an exponential distribution which is confirmed by our results (see figure \ref{fig:distribution}). 

\begin{figure}
\centerline{\epsfig{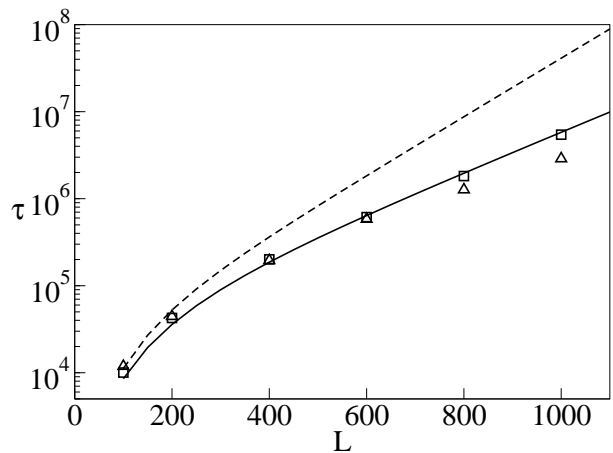}}
\caption{The measured average transition times compared with the theory. The solid/dashed line shows the lifetime of the high-density/low-density state according to the theory as a function of the system size, and the squares/triangles are the results of the MC simulations. Parameters are: $\rho_-=0.2705, \rho_+=0.63, \Omega_a=0.5, \Omega_d=0.1$. For each value of $L$ the number of observed transitions was between 1600 and 2100.}
\label{fig:comparison}
\end{figure}

By measuring the mean first passage times it is possible to check the predictions of section \ref{sec:transitions}. However, it can be carried out only for relatively small systems because of the exponentially increasing time scale. We measured the mean first passage time up to $L=1000$ (see figure \ref{fig:comparison}). The agreement for the passage time from left to right (this is the life time of the HD state) is good, but it seems that the theory gives a wrong result for the mean lifetime of the LD state (note, that there is no particle-hole symmetry in the system). In the following we present a possible explanation for this discrepancy, which we believe to disappear as $L \to \infty$.

For finite systems the hydrodynamic description is not exact, there are finite size effects. Our assumption was that the reaction kinetics do not change the local correlations, but only the value of $\rho$, which becomes $x$-dependent on Euler scale. This is not true in general for finite systems where the reactions modify the correlations and therefore $S(\rho)$ and $j(\rho)$ is changed in (\ref{eq_DE}). Moreover, the effective hopping rates of the shock can also change if there are correlations. In our model, however, there is a special point where the stationary state remains a product measure (the stationary distribution of the ASEP) even for finite systems. The homogenous product state with density $\rho^*\equiv\Omega_a/(\Omega_a+\Omega_d)$ is a stationary distribution of the process (\ref{eq_react}) alone and also together with the ASEP hopping dynamics. Hence it is plausible that the further we are from $\rho^*$ the more pronounced the finite size effects are (note that $\rho(x)=\rho^*$ is always a solution of (\ref{ode})). 

To confirm this we measured the equal time density-density correlations 
\begin{equation}
c_k=\langle n_{L/2} n_{L/2+k}\rangle - \langle n_{L/2} \rangle \langle n_{L/2+k} \rangle
\end{equation}
in the high and low density states ($n_l$ is the occupation number of the $l$th site). The results show that in the HD state $c_k$ is very close to zero (it is zero within the statistical error of our measurement), whereas in the LD state there is a considerable correlation in finite systems. Although the correlation length (on the lattice scale) increases the amplitude decreases with increasing system size.

\begin{figure}
\centerline{\epsfig{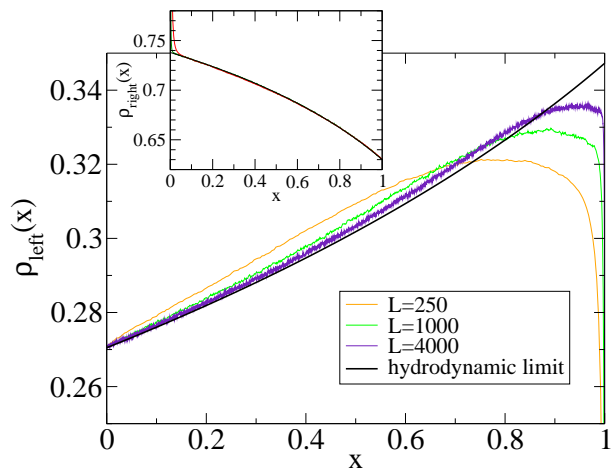}}
\caption{(Color online) Shown is the measured density profile of the low density state for different system sizes compared to the hydrodynamic limit. The average was taken over those configurations of the history where the second class particle was on the last ($L$th) lattice site. It can be seen that the finite size effects are very pronounced even for large systems. For comparison the inset shows the measured density profile of the high density state for $L=250$ and $L=1000$ compared to the hydrodynamic limit. Here, apart from the exponential behaviour at the boundary (which is due to the microscopic structure of the shock) the agreement is perfect even for relatively small systems. Parameters are: $\rho_-=0.2705, \rho_+=0.63, \Omega_a=0.5, \Omega_d=0.1$, which gives $\rho^*=5/6\approx 0.83$.}
\label{fig:rholeft}
\end{figure}

This suggests that the density profile of the high density state is much less sensitive to finite size effects than the one of the low density state. The deviation from the ``true'' $\rho(x)$, by assuming no correlations, increases with increasing/decreasing $x$ in the LD/HD state since they are determined by the left/right boundary density, therefore the agreement with the analytical prediction is the worst in the LD state near the right boundary (see figure \ref{fig:rholeft}). The dominant contribution to the mean first passage time from left to right is determined by that part of the energy landscape which is to the left of the maximum. In this region the density profiles, and therefore also $E(x)$, are less sensitive to finite size effects hence there is a good agreement. On the other hand the energy landscape determined by the hydrodynamic description is much less reliable near the right boundary, which results in a significant difference between the predictions and the measurement of the life time of the LD state for accessible system-sizes (see figure \ref{fig:comparison}). We believe that these are finite size effects and due to the weakening of correlations there is agreement for $L\to\infty$ (compare the range of $L$ in figure \ref{fig:comparison} and figure \ref{fig:rholeft}).  

\begin{figure}
\centerline{\epsfig{file=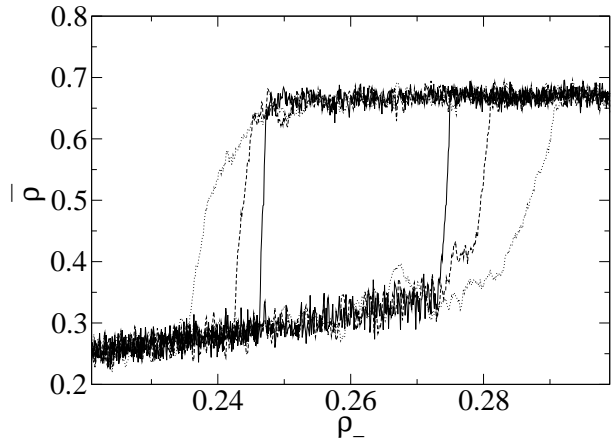,width=7truecm,angle=270}}
\caption{Hysteresis plot for $L=2000, \Omega_a=0.7, \Omega_d=0.1, \rho_+=0.45$. $\rho_-$ was changed by $10^{-4}$ in every 5000 (solid line), 1500 (dashed line), and 500 (dotted line) MC steps (time average was not taken). The hysteresis loop gets wider as the speed of changing $\rho_-$ is increased.}
\label{fig:hysteresis}
\end{figure}

We also found hysteresis \cite{Rako03} by measuring the space averaged density $\bar\rho$ while changing $\rho_-$ in such a way that the system starting from the LD phase passed through the non-ergodic phase and ended up in the HD phase. Then the process of changing $\rho_-$ was reversed. Results are shown in figure \ref{fig:hysteresis}. We found that the hysteresis loop inflates with increasing speed of sweeping, which is reminiscent of hysteresis in usual magnetic systems.

\section{Conclusions}
\label{sec:conclusions}

We found a class of noisy one-dimensional systems with only one species of particles which can show ergodicity breaking in the thermodynamic limit. These systems can be constructed by taking a conserving one-species driven diffusive system on an open one-dimensional lattice and introducing bulk non-conservative reaction-kinetics with rates proportional to the inverse system size. A general property of these systems is that the stationary density profile is not constant in the bulk, moreover it is often non-continuous, i.e., a shock is localized at a macroscopic distance of the boundaries.

By using a hydrodynamic description on the Euler-scale we find the dominant dynamical mode, which is the random walk of a shock in an effective potential (within a confined volume). It is surprising that in these non-equilibrium many-body systems the major role is played by a single-particle collective mode moving under equilibrium conditions. In the above framework localized shocks correspond to local minima of the potential, while a local maximum generates ergodicity breaking, since the effective energy barrier is proportional to the system size $L$. Based on this physical picture we give a criterion for the existence of a non-ergodic phase. In this phase there are more stationary states all of which correspond to a local minimum of the potential. In finite systems transitions between these states are possible but the average transition time is exponentially large in the system-size. 

We study one particular model in detail which is the ASEP with the $A\varnothing A\rightleftharpoons AAA$ reaction kinetics \cite{Rako03}. With the moving shock in mind it is possible to derive the phase diagram  of the system analytically, which indeed exhibits a phase with broken ergodicity. We also performed Monte Carlo simulations to confirm our findings and the results are in good qualitative and quantitative agreement with the theory apart from the average transition time from the low density to the high density state in the non-ergodic regime. We argue that this is due to strong finite size effects. Unfortunately the accessible system sizes are very much limited in these simulations because of the exponentially growing time scale. An analytical calculation of correlations present in finite systems would be needed for a better understanding of this discrepancy. 

Ergodicity breaking is often associated with spontaneous symmetry breaking (SSB) but we would like to emphasize that the former is much more general. In the systems we consider, usually there is no symmetry breaking since there is no symmetry which would relate the different stationary states to each-other. However, in special cases the ergodicity breaking we observe does reduce to SSB. One example is the ASEP with the reaction kinetics $A\varnothing A\to AAA$ and $\varnothing A\varnothing\to \varnothing\varnothing\varnothing$ . If these reactions have the same rate $\Omega$ and the two boundary densities are related by $\rho_+=1-\rho_-$ then the exchange of particles and holes followed by a space reflection is a symmetry. According to (\ref{Source}) the source term of this model is 
\begin{equation}
S(\rho)=-\Omega\rho(1-\rho)(1-2\rho),
\end{equation}
which satisfies the criterion for ergodicity breaking (see sec. \ref{sec:effective_potential}). The two solutions of (\ref{eq_DE}) fitting the left and right boundary densities are
\begin{eqnarray}
\label{SSBdensity1}
\rho_\mathrm{left}(x)=\left( 1+\left( \frac{1}{\rho_-}-1\right)e^{\Omega x} \right)^{-1}, \\
\label{SSBdensity2}
\rho_\mathrm{right}(x)=\left( 1+\left( \frac{1}{\rho_+}-1\right)e^{\Omega (x-1)} \right)^{-1}.
\end{eqnarray}
These density profiles are related by the above symmetry and generate an effective potential (\ref{hopping_rates},\ref{potential}) which is symmetric and has a single maximum at $x=1/2$. For $\rho_-=1-\rho_+<1/2$ the theory predicts two stationary states which correspond to the shock being at the left and right boundaries and are characterised by the above density profiles (\ref{SSBdensity1},\ref{SSBdensity2}) \footnote{For $\rho_-=1-\rho_+>1/2$ the boundary densities have to be replaced by $1/2$ in equations (\ref{SSBdensity1},\ref{SSBdensity2}) because of a boundary layer which is similar to that of the ASEP in the maximal current phase.}. 

This is a novel kind of mechanism for SSB as compared to the bridge model \cite{Evan95}, where the SSB is triggered by blockages at the boundaries. These blockages can be considered as effective impurities \cite{Popk04a} and induce the breakdown of the hydrodynamic description by inhibiting the identification of boundary densities in a consistent way. Once these blockages are removed the system becomes ergodic and no symmetry breaking is possible. In contrast, in the systems we consider there are no such impurities. The boundary densities are well-defined and the hydrodynamic description works. However, in some cases the hydrodynamic equation has several stable stationary solutions. The dynamical picture behind is the random motion of a shock in an effective potential, which has several local minima.

\section*{Acknowledgments}
We are grateful to G.~M.~Sch\"utz for fruitful discussions and useful suggestions. We thank R.~J.~Harris and D.~Mukamel for valuable comments. A.~R\'akos acknowledges financial support from the Deutsche Forschungsgemeinschaft (DFG).

\section*{References}
\bibliographystyle{unsrt}
\bibliography{references}

\end{document}